\documentclass[]{aa} 

\usepackage{graphics,txfonts}  
\usepackage{subfigure} 
\usepackage{rotating}

\newcommand{\kms}{km\,s$^{-1}$}

\newcommand{\vs}{${\rm v}_{\mathrm{e}}\sin i$}

\def\i{\,{\sc i}} \def\ii{\,{\sc ii}} 
\def\hd{HD\,50773}
\def\co{CoRoT}
\def\llm{{\sc LLmodels}}

\begin{document}

\title{Surface structure of the CoRoT\thanks{The \co\  space
mission was developed and is operated by the French space agency CNES, with
participation of ESA's RSSD and Science Pograms, Austria, Belgium, Brazil,
Germany, and Spain.} CP2 target star \hd}

\author{T. L\"uftinger\inst{1}, H.-E. Fr\"ohlich\inst{2}, W. Weiss\inst{1}, P. Petit\inst{3}, M. Auri\'ere\inst{3},
N. Nesvacil\inst{1}, M.~Gruberbauer\inst{1}, D. Shulyak\inst{1},
E.~Alecian\inst{4}, A. Baglin\inst{4}, F. Baudin\inst{7}, C. Catala\inst{4},
J.-F.~Donati\inst{3}, O. Kochukhov\inst{5}, E. Michel\inst{4,6}, N.
Piskunov\inst{5}, T.~Roudier\inst{3}, R.~Samadi\inst{4,6} 
        }

\titlerunning{The CoRoT CP Target Star \hd}
\authorrunning{T. L\"uftinger et al.}
\offprints{~\\ \email{lueftinger@astro.univie.ac.at}}
\institute{Institut f\"ur Astronomie, Universit\"at Wien, T\"urkenschanzstrasse 17, 1180 Wien, Austria
      \and Astrophysikalisches Institut Potsdam, An der Sternwarte 16, D-14482 Potsdam, Germany
      \and Laboratoire d'Astrophysique de Toulouse-Tarbes, Université de Toulouse, CNRS, France  
      \and Observatoire de Paris, LESIA, 5 place Jules Janssen, F-92195 Meudon Cedex, France
      \and Department of Physics and Astronomy, Uppsala University, 75120 Uppsala, Sweden  
      \and Universit\'{e} Pierre et Marie Curie, Universit\'{e} Denis Diderot, Pl. J. Janssen, 92195 Meudon, France
      \and Institut d'Astrophysique Spatiale, UMR8617, Universit\'{e} Paris X, B\^{a}t. 121, 91405 Orsay, France
            }

\date{Received / Accepted }
\abstract {} 
{We compare surface maps of the chemically peculiar star \hd\ produced with a
Bayesian technique and based on high quality CoRoT photometry with those derived
from rotation phase resolved spectropolarimetry. The goal is to investigate the
correlation of surface brightness with surface chemical abundance distribution
and the stellar magnetic surface field.} 
{The rotational period of the star was determined from a nearly 60 day long
continuous light curve obtained during the initial run of CoRoT. Using a
Bayesian approach to star-spot modelling, which in this work is applied for
the first time for the photometric mapping of a CP star, we derived longitudes, latitudes and
radii of four different spot areas. Additional parameters like
stellar inclination 
and the spot's intensities were also determined. The CoRoT observations
triggered an extensive ground-based 
spectroscopic and spectropolarimetric observing campaign and enabled us to
obtain 19 different high resolution spectra
 in Stokes parameters I and V with NARVAL, ESPaDOnS, and SemelPol spectropolarimeters.
Doppler and Magnetic Doppler imaging techniques allowed us to derive the magnetic
field geometry of the star and the surface abundance distributions of Mg, Si, Ca, Ti,
Cr, Fe, Ni, Y, and Cu.} 
{We find a dominant dipolar structure of the surface magnetic field. The \co\ light curve 
variations and abundances of most elements mapped are
correlated with 
the aforementioned geometry: Cr, Fe, and Si are enhanced around the magnetic
poles and coincide with the bright regions on the surface of \hd\ as predicted by our light curve synthesis 
and confirmed by photometric imaging.} 
{} 

\maketitle

\section{Introduction}

\label{intro} 
CoRoT (Convection, Rotation and planetary Transits) is a space mission with the participation of ESA's Science 
Program and Research and Scientific Support Department (RSSD), Austria, Belgium, Brazil, Germany, and Spain. It focuses on high precision photometry from 
space, also taking advantage of observing given targets continuously during
nearly half a year, which is impossible 
from the ground. A technical overview is presented in Boisnard \& Auvergne (2006), and the asteroseismology 
related mission aspects are discussed in Baglin et al. (2006) and references therein. 

\hd\ (BD -00~1488, TYC 4801-2-1, mag(B)\,=\,9.50) was observed during the
initial run of \co\ only little more 
than a month after the launch on December 27, 2006, on board of a Soyuz Fregat II-1b. 
The star was chosen in the seismology field as one of ten possible targets because of its location in the classical
instability strip and of its suspected chemical peculiarity. The reduction of the \co\ photometry to the N2 data 
format is described in Appourchaux et al. (2008), and 
our additional reduction steps are given in Sect.\,\ref{obser} of this paper. 

Not much has been published about this star, which is classified in {\sc simbad} 
as an A2 star and as an A4 -- A9 suspected chemically peculiar star in Renson et al.
(1991) with so far no measured magnetic field nor rotation 
period. 

Magnetic Ap stars, to which \hd\ belongs, represent about 1\,\% to 5\,\% of the upper main sequence
stars and exhibit highly ordered, very stable and often very strong magnetic
fields. They frequently show both brightness- and spectral line profile
variations synchronised to stellar
rotation. These variations are hitherto most successfully explained by the 
oblique rotator model (introduced by Stibbs \cite{stibbs1950}) and are attributed to 
oblique magnetic and rotation axes and to the presence of a non-uniform distribution of
chemical elements on their surface. 

The abundance inhomogeneities in turn are said to arise from the selective diffusion
of ions under the competitive action of radiative acceleration and gravitational
settling (Michaud \cite{michaud1970}) under the influence of the (oblique) magnetic field, 
possibly in combination with a weak stellar wind (see, e.g., Babel 1992).

\begin{figure}
\begin{center}
\includegraphics[width=0.5\textwidth]{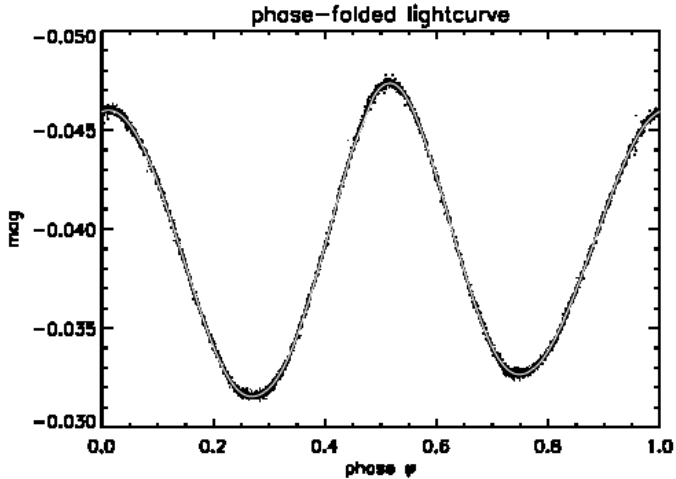}
\caption{Phase plot of \hd\ using a binned light curve after the subtraction 
of effects due to the \co\ orbit and the stellar signal not attributed to rotation. 
Overplotted (white line) is the four-spot model fit described in
Sect.\,\ref{Bayes} of this Paper.}
\label{fig:phase_folded}
\end{center}
\end{figure}

The light curve of \hd\ obtained by CoRoT has a double wave  
form with two clear maxima of slightly different amplitudes at phases of 
 $\phi\,\approx\,0.05$ and $0.52$ 
(see Fig.~\ref{fig:phase_folded}). This photometric variability is, as mentioned, likely to be connected with the inhomogeneous
surface element distribution as shown in e.g., Krti\v{c}ka et al. (2007)
and references therein. As the star seems to be of the Cr CP2-type, we expect
brighter spots rather than dark ones (see e.g. Mikul\'{a}\v{s}ek~et al. 2007 and
Mikul\'{a}\v{s}ek~et al. 2008).
Bright photometric spots seem to be closely connected with overabundance regions
of Si or Fe, and the mechanism of the origin of their contrast in the optical
spectrum region 
is well described in Krti\v{c}ka~et al. (2009). 

It is still not understood why some of the upper main sequence stars are
magnetic and chemically peculiar and others are not, where the 
magnetic fields of CP2 stars originate from, and how these fields exactly
interplay and correlate with the inhomogeneous surface distribution of chemical elements.  
Deriving in detail the surface structure of \hd\ in the way presented in this paper 
will contribute to solving this puzzle.

Our paper is organised in Sect.\,\ref{obser}, where we present our
\co\ space photometry, spectropolarimetric observations, 
and the data reduction, while the physical parameters of the target star \hd\ and a
detailed abundance analysis are presented in Sect.\,\ref{atmospheric
parameters}. In Sect.\,\ref{Bayes} we discuss our analysis of the surface
structure of \hd\ by applying Bayesian methods to the \co\ light curve. Details about the magnetic field geometry and
the surface abundance structures of individual elements are presented in
Sects.\,\ref{magnetic field geometry} and \ref{abundance mapping}. In
Sect.\,\ref{models} we discuss the theoretical predictions of light
variability, taking into account the surface abundance inhomogeneities, and
Sect.\,\ref{discuss} is devoted to the comparison of our different approaches
and the discussion of results. 
 
\section{Observations and data reduction}
\label{obser}

\subsection{CoRoT Data}

HD 50773 was observed by CoRoT from 
February 3 to April 2, 2007. 
The reduction of the \co\ photometry to the N2 level was performed as described 
in Appourchaux et al. (2008).
The time span of 57.7 days covers 27.6 stellar rotational cycles, and the overall variation in
brightness amounts to 0.017 mag. The wavelength range (in the \co\ seismology
field) defined by the CoRoT photometric CCD 
covers $2500$\AA\ to $11000$\AA. For additonal instrumental details, we kindly 
refer the reader to Fridlund et al. (2006).   

\begin{figure}[htbp]
\begin{center}
\includegraphics[width=0.4\textwidth]{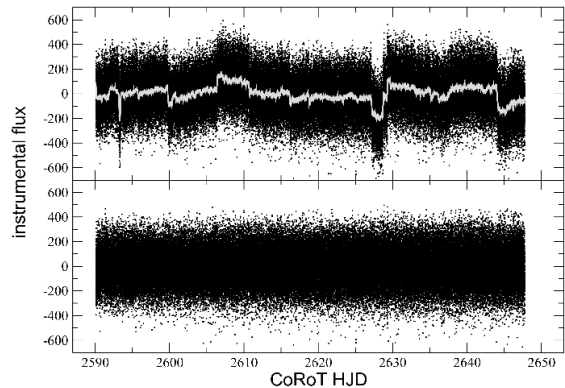}
\caption{\emph{Upper panel:} The CoRoT light curve of HD\,50773 after removal of the obvious stellar variability (black). 
Some small flux changes are remaining, which are shown in grey after
applying a running average (see text in this Sect.).  
\emph{Lower panel:} After the subtraction of this running average, 
the residuals to the stellar variability are mostly consistent with white noise. 
Note that the abscissa in this plot is in \co\,-\,HJD\,(\,=\,HJD\,-\,2451545.0) and the
ordinate in \co\ - internal flux units of the N2 data level after subtraction of
the mean photometric signal, which is 301700 in the same units.}  
\label{fig:noiseresid}
\end{center} 
\end{figure}

Any instrumental signal or stellar variation not associated with rotation and not consistent 
with Gaussian noise or a linear trend would affect the determination of the various parameters 
in our spot model. Therefore, we tested the light curve for any jumps in flux,
which could either be caused by the star itself or the instrument. We first subtracted
a linear trend from the time 
series attributed to aging effects in the detector and electronics chain (Auvergne~et~al.~2009)  
and subsequently removed the apparent stellar rotational variability from the data. Low-order
polynomials were fitted to short subsets $\left(\sim 0.2 \rm\,\right)$ of the residual data in order to eliminate 
the signal changes due to rotation, which resulted in a residual time series
shown in the upper panel of Fig.\,\ref{fig:noiseresid}. In a next step we calculated 
a running average of this residual time series using a box 
width of 370 data points and subtracted the running average - resampled 
to the original time tags - from the original light curve.
The result is shown in the lower panel of Fig.\,\ref{fig:noiseresid}.    

\begin{figure}[htbp]
\begin{center}
\includegraphics[width=0.5\textwidth]{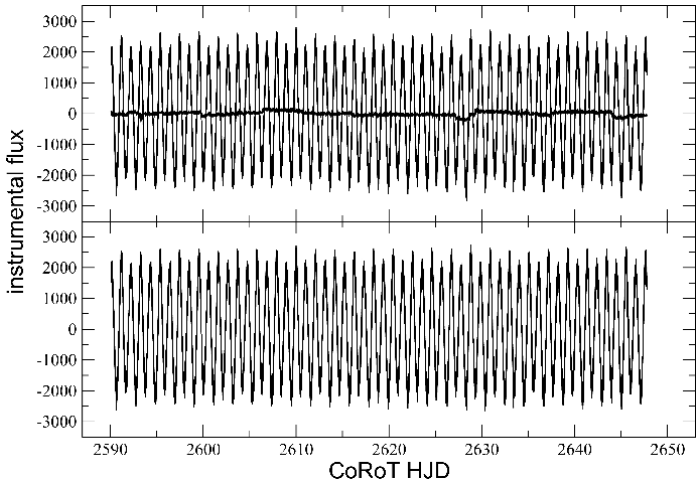}
\caption{\emph{Upper panel:} The detrended CoRoT light curve of HD\,50773 and the resampled running average containing information 
 about residual signal obviously not related to stellar rotation (thick black line). \emph{Lower panel:} The final light curve containing only
signal due to stellar rotation and used for the Bayesian analysis, but zero-mean corrected for the sake of
comparability. Again the abscissa in this plot is in \co~-\,HJD\,(\,=\,HJD - 2451545.0) and the
ordinate in \co\ - internal flux units of the N2 data level after subtraction of
the mean photometric signal, which is 301700 in the same units.} 

\label{fig:corresid}
\end{center}
\end{figure}

The upper panel of Fig.\,\ref{fig:corresid} shows the original \co\ light curve with 
the cumulated corrections obtained so far as a thick black line. After the subtraction of these corrections we 
arrive at the final 
\co\ light curve of \hd\ (lower panel of Fig.\,\ref{fig:corresid}). To reduce the large 
amount of 139271 data points (and the computing time) we binned the data as follows. The un-binned 
autocorrelation function shows a sharp drop within 0.0006 days and goes through
zero at a timelag of 0.02 days. 
As a compromise between high time resolution - at least a
fourth of CoRoT's   
orbital period, i. e. 0.018 days - and statistical independence, a binning length of 
$\approx$\,0.01~days was chosen. Each timebin is represented by the median, and the 
assigned weight is simply the number of data used, which on average were 
26.7 original data points. The binned light curve is used for our analysis described in the 
following 
sections. 

\subsection{Spectropolarimetry with NARVAL, ESPaDoNS and SemelPol}

Spectropolarimetric observations of \hd\ were obtained at the Canada-France-Hawaii-Telescope 
(CFHT) using ESPaDOnS (Donati et al. 2006) and NARVAL, which is attached to the Telescope Bernard Lyot (TBL) 
at Pic du Midi (February 2007, simultaneously to the COROT observations), and in December 2007 with
NARVAL and SemelPol in combination with UCLES at the Anglo-Australian Telescope (AAT).

ESPaDOnS and NARVAL are twin spectropolarimeters, consisting of a Cassegrain polarimetric module and a 
fiber-fed \'{e}chelle spectrometer allowing the whole (polarimetrically analysed) 
spectrum from 
3700\AA\ to 10000\AA\ to be recorded in each exposure. 
ESPaDOnS and NARVAL were used in polarimetric mode with a spectral resolution of
R\,$\approx$\,65000. Stokes I (unpolarised) and Stokes V (circularly polarised)
parameters were obtained by means of four sub-exposures between which the 
retarders (Fresnel rhombs) were rotated in order to exchange the beams 
in the whole instrument and to reduce spurious polarisation signatures. 
The extraction of all spectra was done using Libre-ESpRIT (Donati et al. 1997),
a fully automatic reduction package installed both at CFHT and TBL. 

SemelPol is a visitor instrument, which is mounted at the Cassegrain focus of the
AAT in combination with the UCLES spectrograph.
This combination has been described in detail by e.g. Semel et al. (1993), or
Donati et\,al. (1999 and 2003). In this setup, two beams of opposite polarisation feed
into the \'{e}chelle spectrograph through two separate optical fibres. 
During the observing sequence consisting of four subexposures to obtain Stokes I and V, the
azimuth of the quarter-wave plate 
is switched back and forth
between +45\degr\ (first and fourth exposure) and -45\degr\ (exposures 2 and 3),
allowing for a removal of systematic erros in the measurements. 
Spectra cover the wavelength range between $\approx$\,4300\AA\ and 6800\AA\ with a resolving
power of R\,$\approx$\,70000. The above mentioned Libre-ESpRIT (Donati et al. 1997) package was used for data reduction. 
 
We observed \hd\ during 12 nights and obtained 19 Stokes V (and Stokes I) series
(Table\,\ref{table:1}).

For the Zeeman analysis Least-Squares Deconvolution (LSD, Donati et al. 1997) was applied to 
each observation. To perform the cross-correlation analysis we produced a mask calculated 
for physical parameters as deduced in Sect.\,\ref{atmospheric parameters} and
abundances obtained in Sect.\,\ref{abundances}, using 
spectral line lists from the Vienna Atomic Line Database (VALD; Piskunov et al. 
1995; Kupka et al. 1999; Ryabchikova et al. 1999).
The longitudinal magnetic field (B$_l$) measurements with their 1 $\sigma$ error
bars in G were computed using the first-order moment method (Rees \& Semel 1979, 
Donati et al. 1997).

\begin{table*}
\caption{Journal of the spectroscopic observations of HD 50773: dates,
instrument, exposure time, peak S/N, HJD, phase, B$_l$, $\sigma$.}           
\label{table:1}   
\centering                         
\begin{tabular}{|c|c|c|c|c|c|c|c|}     
\hline\hline               
Date          &Instrument & Exposure time& S/N & HJD        & Phase & $B_{\mathrm{l}}$  & $\sigma$   \\
              &           &              &     &(245 0000+) &       & G      & G          \\
\hline                        
22 Feb. 2007  & NARVAL   &4x600s         & 136 &4154.5095   &0.289  &  50 & 99       \\
28 Feb. 2007  & NARVAL   &4x800s         & 78  &4159.5142   &0.682  & -96  & 223	  \\
28 Feb. 2007  & ESPaDOnS &4x525s         & 130 &4159.7852   &0.812  & -148 & 102       \\
04 Dec. 2007  & NARVAL   &4x600s         & 64  &4439.4471   &0.557  & -179 & 188        \\
05 Dec. 2007  & NARVAL   &4x600s         & 81  &4439.7077   &0.681  & -70  & 161        \\
12 Dec. 2007  & NARVAL   &4x600s         & 156 &4447.4335   &0.376  & 382  & 84         \\
13 Dec. 2007  & NARVAL   &4x600s         & 154 &4447.6877   &0.498  & 441  & 69         \\
13 Dec. 2007  & NARVAL   &4x600s         & 141 &4448.4290   &0.852  &-386 & 94          \\
14 Dec. 2007  & NARVAL   &4x600s         & 123 &4448.6841   &0.974  &-278 & 95          \\
14 Dec. 2007  & NARVAL   &4x600s         & 163 &4449.4622   &0.346  & 219 & 80           \\
15 Dec. 2007  & NARVAL   &4x600s         & 115 &4449.7114   &0.465  & 579 & 98         \\
15 Dec. 2007  & NARVAL   &4x600s         & 119 &4450.4431   &0.815  & -45 & 119        \\
16 Dec. 2007  & NARVAL   &4x600s         & 107 &4450.6943   &0.935  & -382 & 114        \\
16 Dec. 2007  & NARVAL   &4x600s         & 110 &4451.4378   &0.291  & -100 & 138        \\
17 Dec. 2007  & NARVAL   &4x600s         & 106 &4451.6888   &0.411  & 355  & 126        \\
18 Dec. 2007  & NARVAL   &4x600s         & 170 &4453.4605   &0.258  & 9   & 77         \\
19 Dec. 2007  & NARVAL   &4x600s         & 138 &4453.7095   &0.377  & 273 & 94         \\
28 Dec. 2007  & SemelPol &4x2100s        & 170 &4463.1133   &0.875  & -368 & 43         \\
01 Jan. 2008  & SemelPol &4x900s         & 130 &4467.0690   &0.766  & 13  & 64        \\
\hline                                 
\hline
\end{tabular}
\end{table*} 

Rotation phases of \hd\ (Table\,\ref{table:1}) were calculated according to the ephemeris and
rotation period (derived by us): 
 
\hspace{3mm}$HJD = 2454135.09 + 2\fd09101\times E$, where $E$ is an integer
number. 

\section{Atmospheric parameters and abundance analysis}
\label{atmospheric parameters}

\subsection{Atmospheric parameters}
In order to derive accurate atmospheric parameters for \hd, a grid of model atmospheres centered on 
$T_{\rm eff}$= 8500K and log$g$= 4.1 was computed using \llm\
(Shulyak et al. 2004). Synthetic spectra based 
on these models were calculated with synth3 (Kochukhov 2007) and were compared to observations. Atomic parameters used for 
spectrum synthesis were extracted from VALD using 
the default configuration file. Average surface element abundances of spectral lines corresponding 
to different species, which could later be used as starting values for Doppler
imaging, were determined from equivalent 
width measurements and an adapted version of the WIDTH9 code (Kurucz 1993). For this
analysis observations from different phases
were co-added to reduce line profile asymmetries. 
The effective temperature was derived from abundance - excitation potential correlations computed for all model 
atmospheres in the grid using 39 Fe\,I and 15 Fe\,II lines. This correlation is very sensitive to changes in effective 
temperature. For further analysis we adopted a model with $T_{\rm eff}$=
8300\,K, for which the calculated Fe abundance 
was found to be independent of the excitation potential of individual transitions.

For abundance analyses of normal A type stars the surface 
gravity log $g$ is usually determined via the ionisation 
equilibrium of the Fe lines, i.e., for a correct value of log $g$ 
the iron abundance determined from a set of Fe\,I lines is equal 
to the abundance derived from Fe\,II lines. In the atmospheres of 
Ap stars, diffusion processes lead to a vertical stratification of 
Fe, Cr and other elements. Spectral lines corresponding to 
neutral and ionised species of the same element sample different atmospheric layers. 
Therefore abundances determined from these two different line sets will not be 
equal, even for a correct choice of log $g$. For our analysis we decided to use 
the value of log $g$= 4.1 $\pm$0.1, based on Geneva photometry (Burki~et~al.
2009, in prep.).

Due to the rather high \vs\ of \hd, possible vertical abundance stratification within the
stellar atmosphere had to be neglected, which leads to a minimum error
of $\pm$300\,K for $T_{\rm eff}$.  

The smallest slope in the equivalent width-abundance correlation was obtained for microturbulent velocities between 
2 and 3 \kms. This rather large value is possibly due to a combination of the effects of line broadening by the magnetic field 
and the co-addition of spectra from different rotation phases. Therefore,
abundances obtained with $v_{\rm mic}$= 2 \kms\ were used as starting values for
our Doppler imaging. 

\subsection{Abundance analysis}
\label{abundances}
Using the WIDTH9 code and the co-added spectrum of \hd, a crude abundance analysis of 17 species was performed. 
Due to the strong rotational variability of practically all spectral lines in this star, detailed abundances for individual 
rotation phases can only be derived via Doppler imaging. The values given in
this section can therefore be taken only as first order approximations. In
Table\,\ref{tab:abund} we present abundances of 17 species derived from equivalent width measurements in
co-added spectra of \hd, based on $T_{\rm eff}$=8300 K, log$g$= 4.1, and $v_{\rm
mic}$= 0 \kms. For each species we give the number of lines measured and the standard deviations of the derived element abundances. 
A complete list of lines used for this analysis is given online in
Table\,\ref{tab:lines}.
\begin{table}
\caption{Abundances of 17 species derived from equivalent width measurements in
co-added spectra of \hd.}
\label{tab:abund}
\begin{center}
\begin{tabular}{|l|c|c|l|}
\hline \hline
Species & log$N_{\rm elm}/N_{\rm tot}$ & \# of lines & stdev  \\
\hline
C I&	-3.03&	1&	-\\
Mg I&	-4.49&	2&	0.04\\
Si I&	-4.14&	2&	0.22\\
Si II&	-3.93&	2&	0.18\\
Ca I&	-5.60&	1&	-\\
Sc II&	-8.32&	1&	-\\
Ti II&	-6.70&	2&	0.21\\
V II&	-6.74&	1&	-\\
Cr I&	-4.49&	2&	0.33\\
Cr II&	-5.10&	7&	0.32\\
Mn I&	-6.36&	1&	-\\
Fe I&	-4.00&	39&	0.30\\
Fe II&	-3.91&	15&	0.37\\
Ni I&	-5.48&	1&	-\\
Y II&	-9.13&	1&	-\\
Pr III&	-6.57&	1&	-\\
Nd III&	-7.82&	1&	-\\
 \hline \hline
\end{tabular}
\end{center}
\end{table}

\onltab{3}{
\begin{table*}
\caption{Line list used for abundance analysis based on equivalent width measurements in the co-added spectrum of \hd.
Atomic parameters are taken from the VALD database. 
For each line wavelength, excitation potential, and log\,$gf$ are listed.}
\label{tab:lines}
\begin{center}
\begin{tiny}
\begin{tabular}{|l|r|r|r||l|r|r|r|}
\hline \hline
Species & $\lambda$ [\AA] & $E_{\rm low}$ [eV]& log$gf$&species & $\lambda$ [\AA] & $E_{\rm low}$ [eV]& log$gf$\\
\hline
\hline
C I&  6001.1179 &  8.6430 & -2.061	&	Fe I&5445.0424 &  4.3860 & -0.020\\
&&&&						Fe I&5446.9168 &  0.9900 & -1.914\\
Mg I&5183.6040  &  2.7170 & -0.180&		Fe I&5487.7450 &  4.3200 & -0.317\\
Mg I&5528.4050  &  4.3460 & -0.620&		Fe I&5554.8951 &  4.5480 & -0.440\\
&&&&						Fe I&5565.7040 &  4.6080 & -0.213\\
Si I&5645.6130  & 4.9300  &-1.524&		Fe I&5569.6181 &  3.4170 & -0.486\\
Si I&5772.1460  & 5.0820  &-1.358&		Fe I&5572.8424 &  3.3960 & -0.275\\
Si I&5948.5410  & 5.0820  &-0.780&		Fe I&5615.6439 &  3.3320 &  0.050\\
Si II&5055.9840 & 10.0740 &  0.593&		Fe I&5624.5422 &  3.4170 & -0.755\\
Si II&5978.9300 & 10.0740 &  0.004&		Fe I&5633.9465 &  4.9910 & -0.270\\
&&&&						Fe I&5686.5302 &  4.5480 & -0.446\\
Ca I&5581.9650  & 2.5230 & -0.555&		Fe I&5705.9922 &  4.6070 & -0.530\\
Ca I&5588.7490  & 2.5260 &  0.358&		Fe I&5762.9922 &  4.2090 & -0.450\\
Ca I&5594.4620  & 2.5230 &  0.097&		Fe I&5816.3735 &  4.5480 & -0.601\\
Ca I&5857.4510  & 2.9330 &  0.240&		Fe I&5848.1270 &  4.6080 & -1.056\\
&&&&						Fe I&5859.5860 &  4.5490 & -0.419\\
Sc II&5667.1490 &  1.5000 & -1.309&		Fe I&5862.3570 &  4.5490 & -0.127\\
&&&&						Fe I&5914.2010 &  4.6080 & -0.131\\
Ti II&5185.9018  & 1.8930 & -1.490&		Fe I&5930.1799 &  4.6520 & -0.230\\
Ti II&5418.7675  & 1.5820 & -2.000&		Fe I&5934.6549 &  3.9280 & -1.170\\
&&&&						Fe I&6003.0123 &  3.8810 & -1.120\\
V II&5819.9350 &  2.5220 & -1.703&		Fe I&6020.1692 &  4.6070 & -0.270\\
&&&&						Fe I&6024.0580 &  4.5480 & -0.120\\
Cr I&5206.0370 &  0.9410 &  0.019&		Fe I&6027.0509 &  4.0760 & -1.089\\
Cr I&5208.4250 &  0.9410 &  0.158&		Fe I&6056.0047 &  4.7330 & -0.460\\
Cr II&5237.3290 &  4.0730 & -1.350&		Fe II&5004.1950 & 10.2730 &  0.504\\ 
Cr II&5310.6870 &  4.0720 & -2.280&		Fe II&5061.7180 & 10.3080 &  0.284 \\
Cr II&5334.8690 &  4.0720 & -1.826&		Fe II&5169.0330 &  2.8910 & -1.303 \\
Cr II&5407.6040 &  3.8270 & -2.151&		Fe II&5234.6250 &  3.2210 & -2.230 \\
Cr II&5420.9220 &  3.7580 & -2.458&		Fe II&5254.9290 &  3.2300 & -3.336 \\
Cr II&5508.6060 &  4.1560 & -2.252&		Fe II&5316.6150 &  3.1530 & -1.850 \\
Cr II&5620.6310 &  6.4870 & -1.395&		Fe II&5362.8690 &  3.1990 & -2.616 \\
&&&&						Fe II&5427.8260 &  6.7240 & -1.581 \\
Mn I&6021.8190  & 3.0750  & 0.034&		Fe II&5432.9670 &  3.2670 & -3.527 \\
&&&&						Fe II&5439.7070 &  6.7290 & -2.382 \\
Fe I&5065.0180 &  4.2560 &  0.005&		Fe II&5534.8470 &  3.2450 & -2.730 \\
Fe I&5078.9750 &  4.3010 & -0.292&		Fe II&5835.4920 &  5.9110 & -2.702 \\
Fe I&5090.7740 &  4.2560 & -0.400&		Fe II&5952.5100 &  5.9560 & -2.388 \\
Fe I&5133.6885 &  4.1780 &  0.140&		Fe II&5961.7050 & 10.6780 &  0.675 \\
Fe I&5139.4628 &  2.9400 & -0.509&		Fe II&5991.3760 &  3.1530 & -3.540 \\
Fe I&5162.2729 &  4.1780 &  0.020&		&&&\\
Fe I&5202.3360 &  2.1760 & -1.838&		Ni I&5715.0660 &  4.0880 & -0.352\\
Fe I&5232.9403 &  2.9400 & -0.058&		&&&\\
Fe I&5281.7904 &  3.0380 & -0.834&		Y II&5087.4160  & 1.0840 & -0.170\\
Fe I&5353.3736 &  4.1030 & -0.840&		&&&\\
Fe I&5367.4668 &  4.4150 &  0.443&		Pr III&	6053.0044 &  0.0000 & -1.983\\
Fe I&5383.3692 &  4.3120 &  0.645&		&&&\\
Fe I&5393.1676 &  3.2410 & -0.715&		Nd III&5845.0201 &  0.6310 & -1.180\\
Fe I&5434.5238 &  1.0110 & -2.122&		&&&\\

\hline \hline
\end{tabular}
\end{tiny}
\end{center}
\end{table*}

}

\section{Analysing the light curve with Bayesian methods -- Photometric Imaging (PI)}
\label{Bayes} 

Fitting the light curve in terms of a few circular spots needs the estimation
of many parameters. 
The star is described by the inclination angle $i$ and the rotational period $P$. Each spot
needs at least three further parameters: longitude $\lambda$, latitude
$\beta$, and spot radius $\gamma$ (in the following longitude increases in the direction of stellar 
rotation, and the zero point is the central meridian facing the observer at the
beginning of the time series).
Additional parameters are the spot intensity $\kappa$ and 
the coefficient $u$ in the linear limb-darkening law. We use 
$u = 0.4415$ derived within model atmosphere calculations with the ATLAS9 code
(Kurucz 1993) applying the wavelength range from $2500$\AA\ to $11000$\AA, which
is defined by the CoRoT photometric CCD.

Before analysing the data proper prior distributions have to be
assigned. If information on the star's 
inclination $i$ is missing, $\cos i$ is evenly distributed for  
$0 \le i \le \pi/2$. Concerning a latitude $\beta$, the value of 
$\sin\beta$ is assumed to be evenly distributed.  
All non-dimensionless parameters like radii or frequencies 
are represented by their logarithms. This makes certain that the posterior distribution for a 
radius will be consistent with that of an area and likewise the posterior for a
frequency with that of a period, i.\,e. it does not make a difference whether
one prefers radii or areas, frequencies or periods. 
 
The likelihood function is constructed as follows.
Spotted stars which are geometrically similar exhibit the same light
curve, except for an offset in magnitude. This offset is considered
irrelevant and removed by integration. This is even indicated, as
the magnitude of the unspotted star, the zero point, is unknown.
This integration can be done analytically if the measurement
errors are assumed to be Gaussian-distributed over the magnitudes. 

With the $N$ data points $d_i$ gained at times $t_i$, 
their standard deviations $\sigma_i$, the fit
$f_0(t_i)$ and an offset $c_0$, the likelihood is represented by:  
$$\Lambda\left(\sigma_{1\dots N}, c_0, p_1\dots{}p_M; d_i\right) = $$ 
$$\prod_{i=1}^{N}{1\over\sqrt{2\pi}\sigma_i}\exp\left(-{\left(d_i-f_0(t_i, p_j)
- c_0\right)^2\over 2\sigma_i^2}\right).$$ \\
The unknown parameters are denoted by $p_j$, with $j=1\dots{}M,$ {\rm  with}
$M=24$ in the case of four spots.
We set $\sigma_i = s_i\cdot\sigma$, with the relative errors $s_i$ being normalised according to 
$\sum_{i=1}^N{1/s_i^2} = N$.

First integrating analytically the measurement error $\sigma$ -- using
Jeffreys' $1/\sigma$-prior -- and then the uninteresting offset $c_0$, one gets
a likelihood depending only on the spot modelling parameters $p_1\dots{}p_M$.
It is this {\em mean\/} likelihood, averaged appropriately over measurement
error and offset, from which the posterior density distributions for all unknowns 
is obtained by marginalization. This is a great advantage of the Bayesian
approach: It provides not only a set of most probable parameter values but 
moreover {\em expectation\/} values and error bars exclusively from the data.  
We can argue that the method itself determines error and
offset from the data. 

The Markov chain Monte Car\-lo (MCMC) method (cf.\,Press\,et\,al.\,2007) 
has been applied to explore the likelihood mountain in  
a high-dimensional parameter space.
The MCMC technique has already shown its capabilities when analysing
photometric data from the Canadian MOST satellite (Croll 2006, Fr\"ohlich 2007). 

A set of 64 Markov chains was generated. Each chain has to perform some
$10^7$ steps, and after a burn-in period every 
thousandth successful step was recorded in order to suppress the correlation between successive steps. 
Budding's star-spot model (1977) was used to model the light curve. 

\subsection{Results from Bayesian PI}

There are two solutions which fit the light curve equally well: 
one with three ``dark'' spots and another one with four ``bright'' spots. In both
cases the residuals amount to 0.12 mmag. The autocorrelation function of the
residuals reveals that there is even more in the data than what can be
represented by a simple model with circular spots. From a formal point of view the
solution with ``dark'' spots is the more probable one, because the fit needs only three
spots instead of four. But in view of the spectroscopic evidence the four-spot
solution makes much more sense. 

The expectation values with 1\,-\,$\sigma$ confidence limits for
the star and the starspot parameters are presented in Table\,\ref{tab01}. From spot longitudes, 
spot epochs (HJD) can be computed: \\
$E_i = 2454135.09 - 2.09101\cdot \lambda_i/360^\circ$.

The reader should be aware that the estimated parameter values and their 
(surprisingly small) error
bars are those of the model constrained by the data. They make sense
{\em given the model is true\/}, i.e. that there are four circular spots,  
a linear limb-darkening law with prescribed coefficient holds and so forth. 

In a further run the coefficient $u$ of the limb-darkening law has been
determined from the data itself. This does not lead to any improvement of
the fit. The fact that the deduced limb-darkening coefficient 
$u = 0.405^{\ +0.010}_{\ -0.013}$ proves to be only a little bit smaller than the
theoretical value of 0.4415 strengthens the confidence in
the reliability of the four-spot model.

The high accuracy of the photometric data makes it reasonable to question
the common wisdom and look for spot evolution as well as differential rotation and even for a period
drift in an Ap star. 
Time evolution is described by a Legendre expansion of the logarithm of the
spot areas up to the seventh power in time. 
We have done this in the case of the three dark spots only. The result is a clear null
result: we neither find a spot area evolution above a 2.5\,-\,$\sigma$ level 
nor a hint for any departure from rigid rotation. 
The spot periods are constant at $\la$ 3 ppm. The lapping time for the two
largest (``dark'') spots would have to exceed 100 years. 

It is important to clean the data from a CoRoT orbit effect: 
The residuals drop by a factor of five by subtracting a periodic
signal containing CoRoT's orbital period as well as three overtones
(Table\,\ref{tab02}). 

\begin{table}
\centering
\caption{Spot parameters: we list {\em expectation\/} values  
and 1\,-\,$\sigma$ uncertainties, the period $P$ is given in days, the spot intensity is presented in units 
of the intensity of the unspotted surface.}
\label{tab01}
\begin{tabular}{|l|lc|r|l|}
\hline
\hline 
\multicolumn{2}{|l}{parameter} & &\multicolumn{2}{|l|}{expectation and uncertainties} \\
\hline
inclination		&$i$                    & &$ 40^\circ\hspace{-3pt}.3$   &$^{+1.3}_{-1.2}$          \\
1st 	longitude     	&$\lambda_1$            & &$184^\circ\hspace{-3pt}.6$	&$^{+0.4}_{-0.4}$          \\
2nd 	longitude     	&$\lambda_2$      	& &$140^\circ\hspace{-3pt}.2$	&$^{+0.5}_{-0.5}$          \\
3rd 	longitude     	&$\lambda_3$       	& &$ 32^\circ\hspace{-3pt}.7$	&$^{+0.3}_{-0.3}$          \\
4th 	longitude     	&$\lambda_4$       	& &$312^\circ\hspace{-3pt}.8$	&$^{+0.3}_{-0.3}$          \\
1st 	latitude     	&$\beta_1$              & &$ 18^\circ\hspace{-3pt}.5$	&$^{+0.8}_{-1.0}$          \\
2nd 	latitude     	&$\beta_2$      	& &$ 21^\circ\hspace{-3pt}.3$	&$^{+1.0}_{-1.0}$          \\
3rd 	latitude     	&$\beta_3$       	& &$ -7^\circ\hspace{-3pt}.3$	&$^{+0.4}_{-0.4}$          \\
4th 	latitude     	&$\beta_4$       	& &$-20^\circ\hspace{-3pt}.2$	&$^{+1.0}_{-1.1}$          \\
1st 	radius     	&$\gamma_1$       	& &$ 23^\circ\hspace{-3pt}.9$	&$^{+0.5}_{-0.4}$          \\
2nd 	radius       	&$\gamma_2$      	& &$ 15^\circ\hspace{-3pt}.0$	&$^{+0.5}_{-0.5}$          \\
3rd 	radius       	&$\gamma_3$      	& &$ 31^\circ\hspace{-3pt}.3$	&$^{+0.4}_{-0.5}$          \\
4th 	radius       	&$\gamma_4$      	& &$ 39^\circ\hspace{-3pt}.3$	&$^{+0.8}_{-0.8}$          \\
 	period       	&$P$                    & &$2.091007$               	&$^{+0.000007}_{-0.000007}$\\
spot intensity          &$\kappa$               & &$1.22$                       &$^{+0.01}_{-0.02}$        \\
\hline
\hline 
\end{tabular}
\end{table}

\begin{table}
\centering
\caption{CoRoT orbit effect. The estimated CoRoT orbital period is given in min, amplitudes in ppm.}
\label{tab02}
\begin{tabular}{|l@{~}ll|*{1}{r@{}l}|}
\hline \hline
\multicolumn{3}{|l}{parameter} & \multicolumn{2}{|l|}{expectation and uncertainties} \\
\hline
\multicolumn{2}{|l}{CoRoT orbital period}    
                                &$P_{\rm orbit}$ & $103.0660$&$\pm{0.0012}$\\[4pt]
amplitude &($P_{\rm orbit}  $)  &$A_1$           & $111$   &$\pm{2}$       \\[4pt]
amplitude &($P_{\rm orbit}/2$)  &$A_2$           & $ 33$   &$\pm{2}$       \\[4pt]
amplitude &($P_{\rm orbit}/3$)  &$A_3$           & $  8$   &$^{+3}_{-2}$   \\[4pt]
amplitude &($P_{\rm orbit}/4$)  &$A_4$           & $ 13$   &$\pm{2}$       \\[4pt]
\hline
\hline
\end{tabular}
\end{table}

\begin{figure*}[htbp]
\centering
\includegraphics[trim = 0mm 1mm 0mm 8.9cm, clip, width=\textwidth]{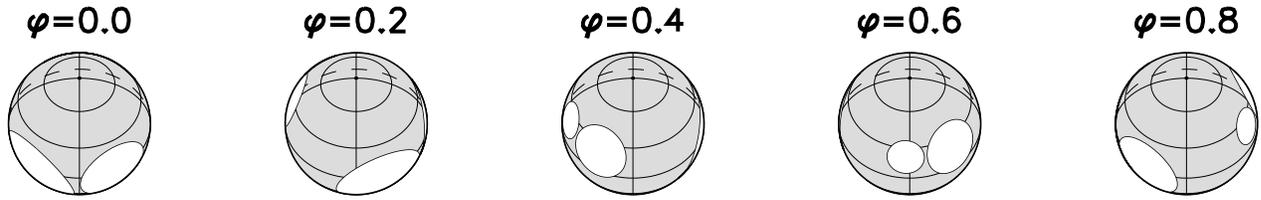}
\caption{Locations of the four bright photometric spots, assumed to be 
of circular shape, plotted at five equidistant phases.} 
\label{photometry spots}
\end{figure*}

\section{Magnetic field geometry}
\label{magnetic field geometry}

All 19 Stokes V LSD-line-profiles derived from ESPaDOnS, NARVAL, and SemelPol spectropolarimetric 
observations are grouped in a single time-series, in order to increase the 
rotational sampling as much as possible and build a proper data set for tomographic inversion. The time-series is 
then used to reconstruct the surface magnetic geometry of the star by means of
Magnetic Doppler imaging (Donati \& Brown 1997).

\begin{figure}[htbp]
\centering
\includegraphics[width=4.4cm]{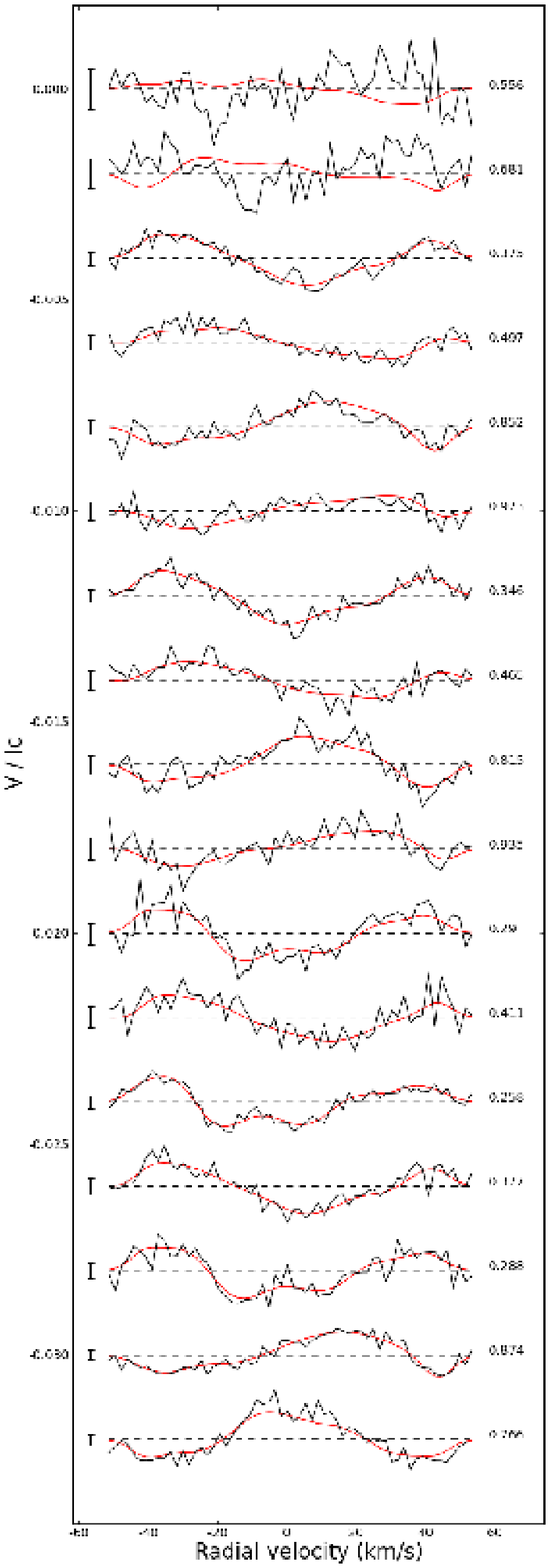}
\caption{Stokes V profiles of \hd\ after the correction of the mean radial velocity of the star. 
Black lines represent the data and red/grey lines correspond to the synthetic profiles of our magnetic model. 
Successive profiles are shifted vertically for better visibility. Rotation phases of observations are 
indicated in the right part of the plot, and error bars are illustrated on the left of each profile.}
\label{fig:stokesv}
\end{figure}

\begin{figure}[htbp]
\centering
\includegraphics[width=9cm]{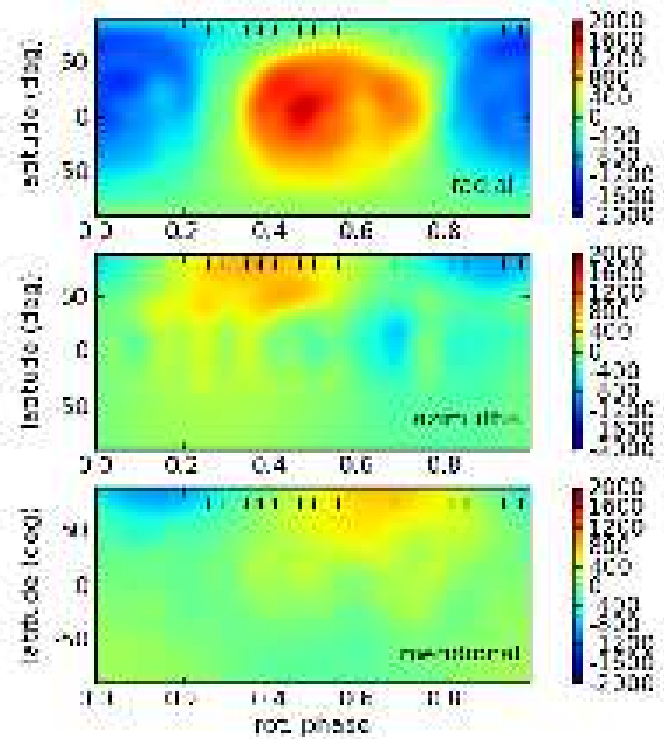}
\caption{Magnetic map of \hd, reconstructed from Magnetic Doppler imaging. Each chart illustrates 
the field projection onto one axis of the spherical coordinate frame with, from top to bottom, the 
radial, azimuthal and meridional field components. The magnetic field strength
is expressed in Gauss, 
and the rotational phases of the observation are indicated as vertical ticks at the top of each chart.}
\label{fig:map}
\end{figure}

To compute synthetic Stokes V line profiles, we make a model of a spherical stellar surface divided into a grid of 
pixels, each pixel producing a local Stokes I and V profile. Assuming a 
given magnetic field strength and orientation for each pixel, local Stokes V profiles are 
calculated under the assumption, valid in the weak field limit, that Stokes V is proportional 
to $g\cdot \lambda_0^2 \cdot B_\parallel \cdot \partial I / \partial \lambda$, where $\lambda_0$ 
is the average wavelength of the LSD profile (about 534 nm for \hd), $B_\parallel$ 
is the line-of-sight projection of the local magnetic field vector, $g$ is the effective 
Land\'{e} factor of the LSD profile (equal to 1.2) and $\partial I / \partial \lambda$ the 
wavelength derivative of the local Stokes I line profile. We further assume that there 
are no large-scale brightness or abundance inhomogeneities over the stellar surface, so that synthetic 
Stokes I profiles are locally the same over the whole photosphere. However we would
like to mention that some features in the Stokes V profile variability 
may arise from the influence of chemical spots and not from the magnetic
topology itself, which has been demonstrated in Kochukhov~et~al.~(2002). 

For each pixel, 
we impose a Gaussian shape to the local synthetic Stokes I line profile. Parameters of the local 
profile (width and depth) are chosen to optimise the adjustment of Stokes\,I LSD profiles of HD 50773.
The amplitude of the 
local Stokes profiles for pixels located at the visible hemisphere of the star is then weighted 
according to a linear limb-darkening coefficient equal to 0.49, and their central wavelengths are 
shifted according to the line-of-sight velocities of individual pixels, assuming
$v\sin i$ = 46.3 \kms\ 
(a value minimising the $\chi^2$ of the magnetic model) and $i=40$\degr.

Synthetic Stokes V profiles are computed for the 19 observed rotation phases
and are compared to the 
observations. The data adjustment is iterative and based on a maximum entropy algorithm 
(Skilling \& Bryan 1984). The version of the code used here makes a projection of the surface magnetic 
field onto a spherical harmonics frame (Donati et al. 2006), with the magnetic field geometry splitted 
between a poloidal and toroidal component (Chandrasekhar 1961). We limit the spherical harmonics 
expansion to $\ell<10$. Using this procedure, the spectropolarimetric data are adjusted at a 
reduced $\chi^2 = 1.2$ (Fig. \ref{fig:stokesv}). The final $\chi^2$, as well as the 
reconstructed magnetic topology, are left essentially unchanged as long as the adopted spherical 
harmonics description respects $\ell_{\rm max} \ge 6$. The reconstructed magnetic geometry is 
illustrated in Fig. \ref{fig:map}, from which we can easily see that the surface magnetic field 
of \hd\ is mostly a highly inclined dipole, with a polar strength of about 2\,kG.  

A closer look at the weight of the complex spherical harmonics coefficients  
$\alpha_{\ell,m}$,  $\beta_{\ell,m}$, and  $\gamma_{\ell,m}$ (defined by Donati et al. 2006) 
is however necessary to obtain a more quantitative information on the magnetic field distribution. 
Whereas the inversion procedure allows for the reconstruction of both poloidal and toroidal 
components of the field, 96\% of the reconstructed magnetic energy ends up in the poloidal 
component. As expected for such a marginal amount of toroidal field, the outcome of the inversion code 
is actually almost similar if we impose the more restrictive case of a purely poloidal field. 
As can be readily seen on the map, most of the magnetic energy (91\,\%) is stored in the dipolar component. 
Higher-order field components are however necessary to obtain a convincing
modelling of the observations 
(up to $\ell = 6$). Finally, the predominance of a highly non-axisymmetric field distribution translates 
into a low amount of magnetic energy in axis-symmetric spherical harmonics modes ($m = 0$), which do not 
gather more than 2\% of the overall photospheric magnetic energy.

\section{Spectroscopic Doppler imaging}
\label{abundance mapping}

By applying the \emph{Doppler imaging} technique (DI), we are able to translate the
partially very pronounced variations in the spectral line profiles of \hd, linked to stellar rotation, into surface maps of the
abundance distribution. 
The longitude of a spot is directly
deduced from the wavelength position of the distortion within the profile, whereas
its latitude can only be derived from time-series observations. In the case of
\hd, 19 such spectra (listed in Table\,\ref{table:1})
could be used for the inversions with {\sc INVERS12}, the DI code we used 
(Kochukhov~et~al.~2004b). 
In this code, where specific intensities are calculated for each visible surface element 
for each iteration, it is possible to simulataneously calculate abundance maps of several chemical elements 
even from blended spectral lines. 
Mapping \hd, we could derive surface abundance distributions for Mg, Si, Ca, Ti,
Cr, Fe, Ni, Y and Cu. We have to mention that Ti and Ni were derived from
spectral line blends where these elements contribute only moderately to the
overall absorption. Input parameters were determined as described in
Sect.\,\ref{atmospheric parameters}, \vs\ and inclination $i$ used as in
Sect.\,\ref{magnetic field geometry}.  

\begin{table*}
\caption{Table of elements and  spectral lines used for mapping of \hd: element, spectral
lines used, the log\,{\it gf} values of these lines, abundance interval in dex, for
comparison the solar values (Asplund~et~al. 2005) are presented. Atomic parameters used
in our study were extracted from VALD.}
\label{lines}
\begin{center}
\begin{tabular}{|l|lrr|c|c|c|} 
\hline
\hline
Species & $\lambda$ (\AA) & log\it gf &  $E_{\rm low}$ (eV) & min, max & blended with & $\odot$  \vspace{-1.7mm}\\
        &                 &           &                     & $log(N/N_{\mathrm{tot}})$ &   &  $log(N/N_{\mathrm{tot}})$ \\
\hline 
\hline 
Mg \i\      & 5401.5210  &  -0.340   & 11.6300  &    		    & Fe, Y  & -4.51 \\
\hline 
Si \ii\     & 5055.9840  &   0.593   & 10.0740  &  $-4.8, -2.6$   &	     & -4.53 \\
Si \i\      & 5421.1680  &  -2.250   &  5.6190  &  $  	 $  & Cr, Mn & \\
Si \i\      & 5421.3830  &  -1.480   &  5.6190  &  $  	 $  & Cr, Mn & \\
\hline 
Ca \i\      & 6102.7230  &  -0.862  &   1.8790  &  $-6.5, -2.7$   & Fe     & -5.73 \\
Ca \ii\     & 6456.8750  &  -0.539  &   8.4380  &  		    & Fe     &       \\
\hline 
Ti \ii\     & 5154.0680  & -1.750   &   1.5660 &  $-8.3, -6.1$   & Cr, Fe & -7.14 \\
\hline 
Cr \ii\     & 5237.3290  & -1.350   &   4.0730 &  $-7.1, -3.6$   &	     & -6.40 \\
            & 5280.0540  & -2.316   &   4.0740 &  		    &	     &       \\
            & 5510.7020  & -2.614   &   3.8270 &  		    &  Fe, Y, Ni     &       \\
\hline 
Fe \i\      & 5383.3692  &  0.645   &   4.3120 &  $ 5.4, -3.6 $  &	     & -4.59 \\
            & 5400.5022  & -0.160   &   4.3710 &			 &		     &	   \\ 
	    & 5400.6560  & -2.482   &	3.6350 &			 &			&     \\    
	    & 5401.2689  & -1.920   &	4.3200 &			 &			&     \\   
\hline      
Ni \i\      & 5510.0030   & -0.900   &  3.8470 &  $-6.1, -5.6 $  & Cr, Fe, Y  & -5.81 \\
\hline        
Y \ii\      & 5509.8950   & -1.010   &  0.9920 &  $-9.9, -7.5 $  & Cr, Fe, Ni & -9.83  \\
            & 5662.9250   &  0.160   &  1.9440 &  		    & Fe	 &	  \\ 
\hline
Cu \ii\     & 5153.2300   &  0.217   &  3.7860 &  $-8.0, -6.9 $  & Cr, Ti, Fe & -7.83  \\
\hline
\hline 
\end{tabular}
\end{center}
\end{table*}
\subsection{Surface abundance structures of individual elements}
\begin{figure*}[htbp]
\includegraphics[trim = 13.0mm 34mm 8mm 30mm, clip, width=125mm]{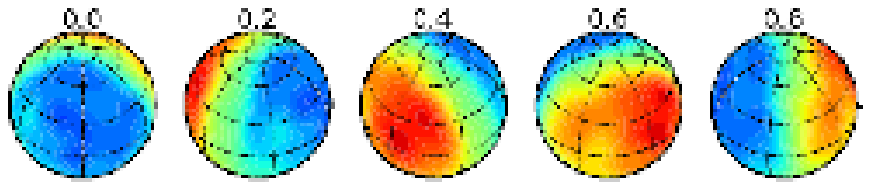}\\
\subfigure 
{
    \includegraphics[trim = 13mm 7mm 146mm 100.33mm, clip, width=21mm]{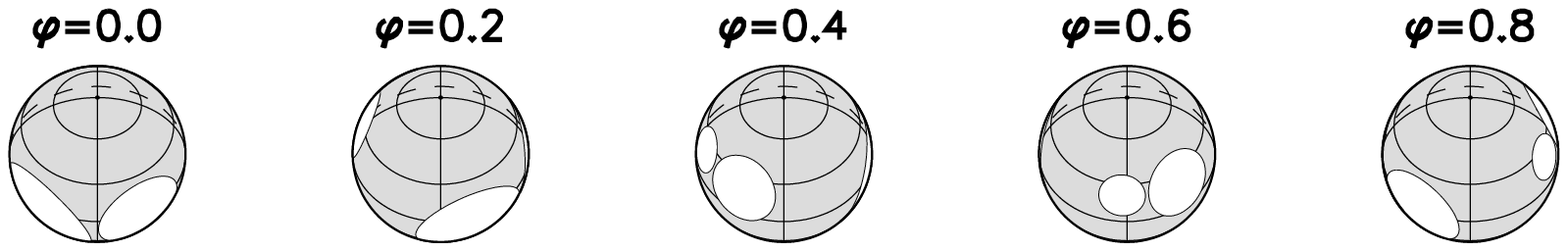}
}
\hspace{0.06cm}
\subfigure 
{
    \includegraphics[trim = 48mm 7mm 111mm 100.33mm, clip, width=21mm]{12239fb7.ps}
}
\hspace{0.06cm}
\subfigure 
{
    \includegraphics[trim = 84mm 7mm 75mm 100.33mm, clip, width=21mm]{12239fb7.ps}
}
\hspace{0.06cm}
\subfigure 
{
    \includegraphics[trim = 120mm 7mm 39mm 100.33mm, clip, width=21mm]{12239fb7.ps}
}
\hspace{0.06cm}
\subfigure 
{
    \includegraphics[trim = 155mm 7mm 4mm 100.33mm, clip, width=21mm]{12239fb7.ps}
}

\includegraphics[width=130mm]{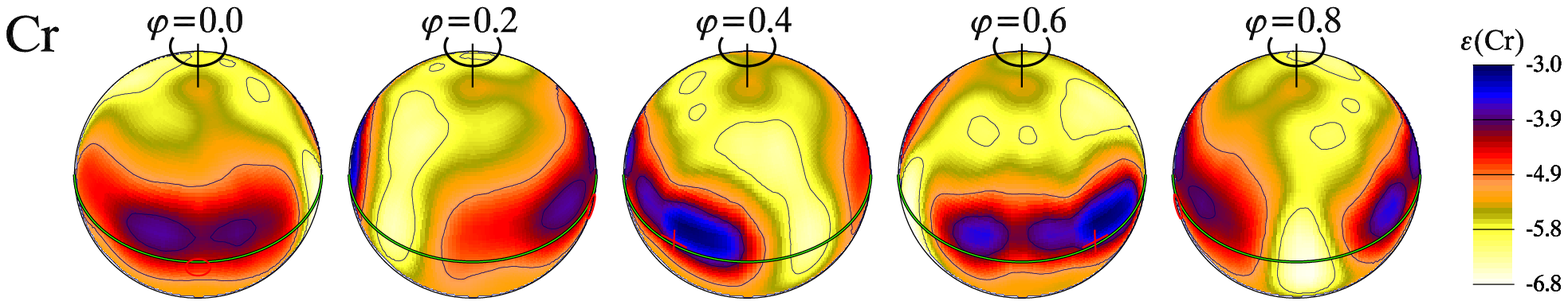}\\
\includegraphics[width=130mm]{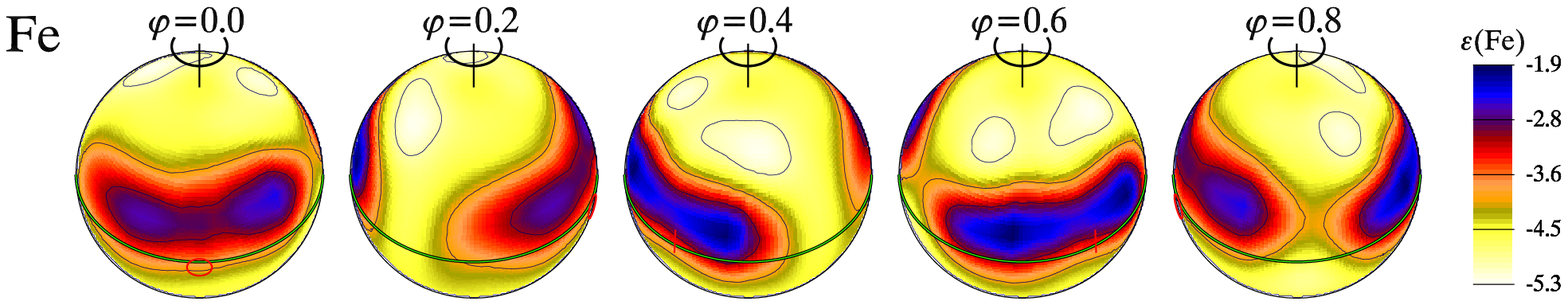}\\
\includegraphics[width=130mm]{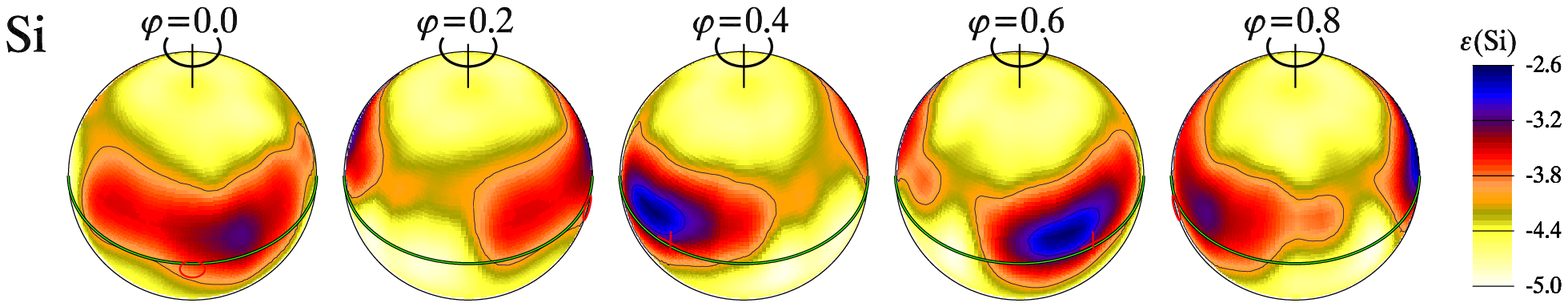}
\label{CrFeSi_maps}
\caption{Top panel: Radial field component of the magnetic map of \hd\ (as described in
Sect.\,\ref{magnetic field geometry}). 
Second panel: locations of the four bright (displayed in black) photometric spots, assumed to be 
of circular shape.  
Next three panels: Abundance distribution of Cr, Fe, and Si at the surface of \hd\ obtained from the lines listed in
Table\,\ref{lines} (online). We show the star at an 
inclination $i$\,=\,40\degr. Darker areas in the plots correspond to higher elemental
abundances, the corresponding scale is given to the right of each panel, and the
contours of equal abundance are plotted with steps of 1.0\,dex. The circle and
the cross indicate the position of the negative and the positive magnetic pole, respectively.
All projections are plotted at five
equidistant rotation phases.} 

\label{maps}
\end{figure*}

\begin{figure*}[htbp]
\begin{center}
\label{MgCaTiNiYCu}
\vspace{8.0mm}
\includegraphics[width=130mm]{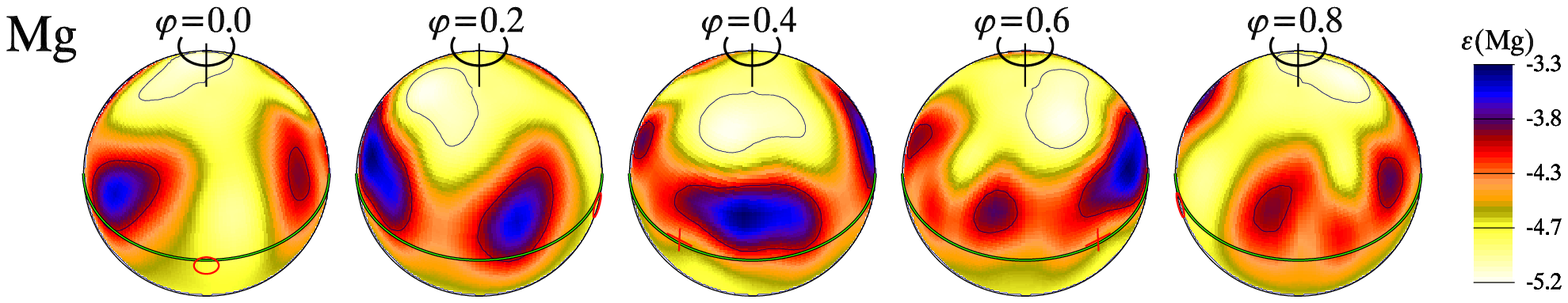}
\includegraphics[width=130mm]{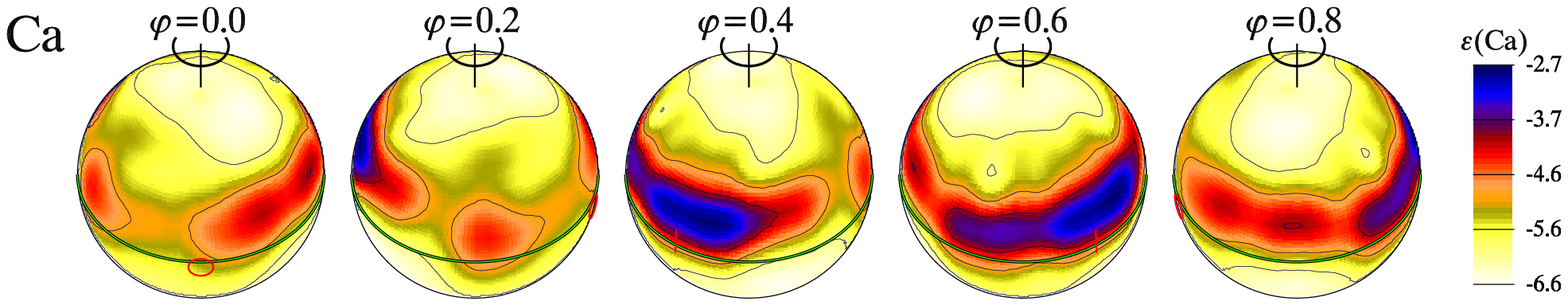}
\includegraphics[width=130mm]{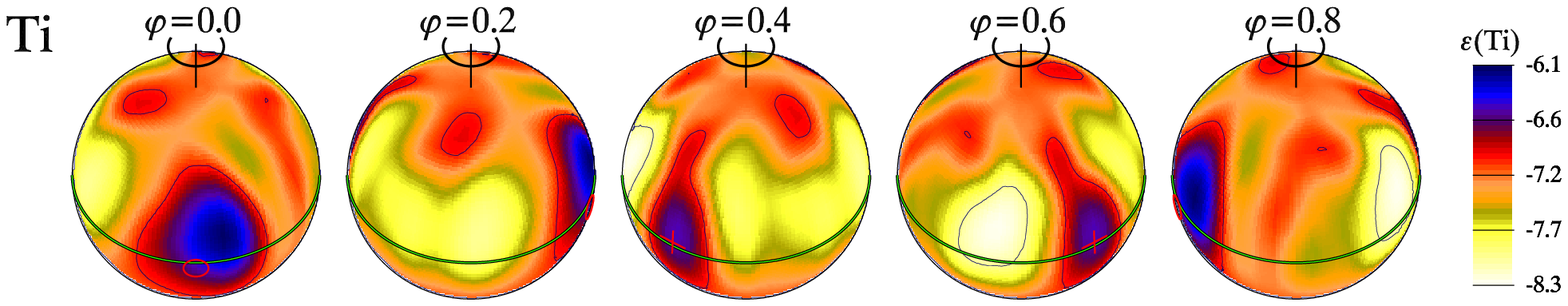}
\includegraphics[width=130mm]{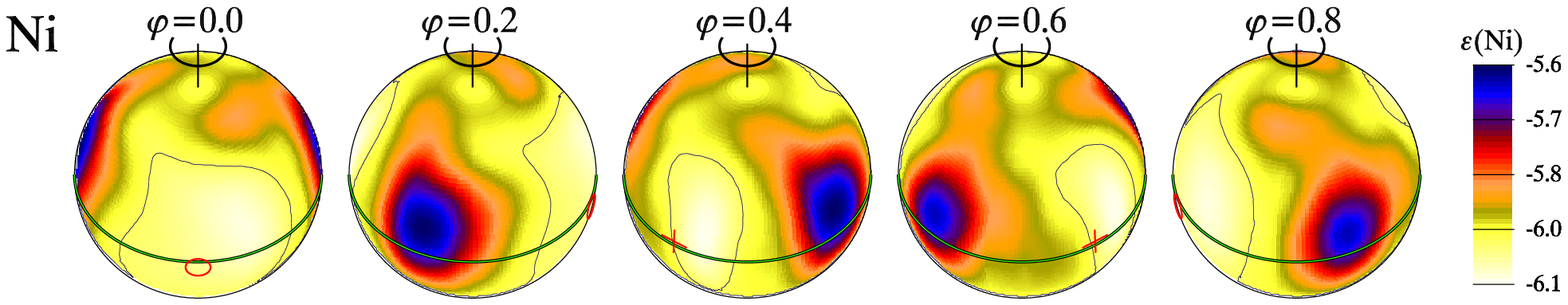}
\includegraphics[width=130mm]{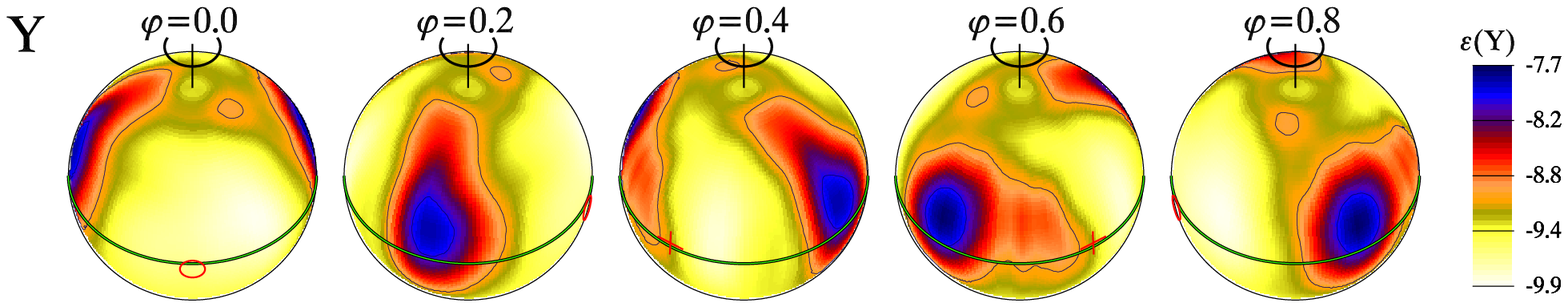}
\includegraphics[width=130mm]{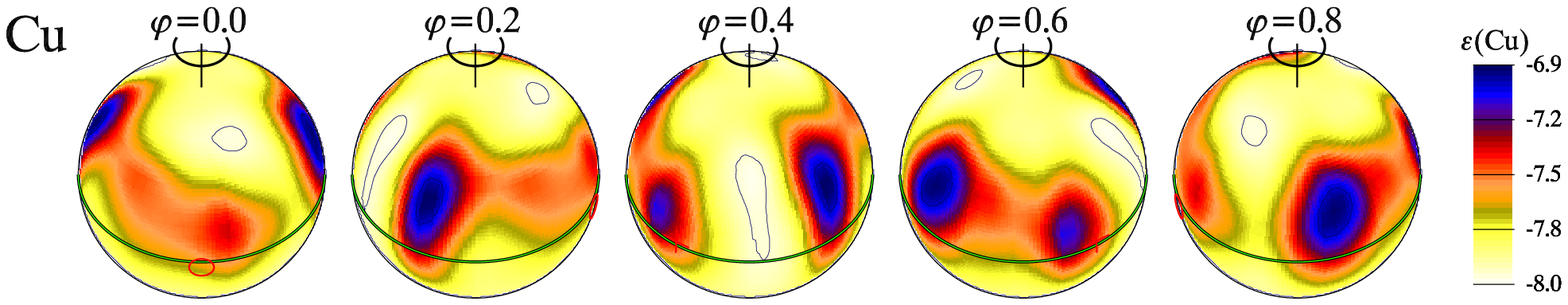}

\caption{Same as the three bottom panels in Fig.\,7 but for Mg, Ca, Ti, Ni, Y, and Cu.} 

\label{maps}
\end{center}
\end{figure*}

\subsubsection{Magnesium and calcium}
The surface abundance distribution of \emph{magnesium} was modeled using a
blend of Mg\i\ with Fe\i\ and Y\ii\ around 5400\,\AA. We find a variation
between $\log N_{\mathrm{Mg}}/N_{\mathrm{tot}}=-5.2$ and $-3.3$\,dex 
and the element, contrary to the other mapped species, does not
exhibit a clear correlation with the magnetic field geometry, and is enhanced in
a spotted, belt-like structure 
near the rotational equator (Fig.\,\ref{MgCaTiNiYCu}). The abundance
distribution of \emph{calcium}, variable between $\log
N_{\mathrm{Ca}}/N_{\mathrm{tot}}=-6.6$ and $-2.7$\,dex and determined from Ca\i\
and \ii\ blended with Fe\i\ around 6102\,\AA\ and 6456\,\AA\ also shows a belt-like structure 
at the rotational equator with a pronounced region of overabundance at the
positive magnetic pole.  

\subsubsection{Iron-peak elements} 
\label{Fe-peak}
A blend of \emph{titanium}\ii\ with Fe\i\ and Cr\ii\ at 5169\,\AA\ was used to recover the surface abundance
structure of this element. Ti\ii\ shows a tendency to accumulate at the magnetic poles and to avoid the magnetic 
equatorial regions, 
varying between $\log N_{\mathrm{Ti}}/N_{\mathrm{tot}}=-8.3$ and $-6.1$\,dex. 

\emph{Chromium} was mapped using several single Cr lines at 5237\,\AA\ and 5280\,\AA\, plus a blend of 
Cr\ii\ with Fe\i\, Y\ii\, and Ni\i\ around 5510\,\AA\ and was found to be variable
between $\log N_{\mathrm{Cr}}/N_{\mathrm{tot}}=-6.8$ and
$-3.0$\,dex. The abundance distribution of the element shows as that of Fe a clear
correlation to the magnetic field geometry, being enhanced at the
poles and depleted at the magnetic equator, whereby the two enhancement regions seem to be subdivided into two spots.  
   
The surface abundance of \emph{iron} was determined using single Fe\i\ lines at
5383\,\AA\, and 5400\,\AA\ and again the blend of Fe\i\ with Cr\ii, Y\ii, and
Ni\i\ at 5510\,\AA. 
As mentioned above, Fe is, as Cr, closely correlated with the magnetic field,
avoiding the magnetic eqatorial region.  
The elemental abundance varies
between $\log N_{\mathrm{Fe}}/N_{\mathrm{tot}}=-5.3$ and $-1.9$\,dex.  

\emph{Nickel}, derived from the blend of this element with Cr\ii, Fe\i, and
Y\ii\ at 5510\,\AA\ and variable between $\log
N_{\mathrm{Ni}}/N_{\mathrm{tot}}=-6.1$ and $-5.6$\,dex shows an opposite correlation 
to the magnetic field geometry from what we saw from Cr and Fe: it is enhanced at the magnetic equatorial 
regions and depleted where the positive and the negative magnetic pole cross the line of
sight.

\subsubsection{Yttrium and copper}
A similar distribution as that for Ni can be observed for \emph{yttrium}. Varying from 
$\log N_{\mathrm{Y}}/N_{\mathrm{tot}}=-9.9$ to $-7.7$\,dex it also shows a pronounced region of
overabundance at the magnetic equator and depletion on the poles. 
The element was mapped from a blend with Cr\ii, Fe\i, and Ni\i\ at 5510\,\AA\ and 5662\,\AA.
\emph{Copper} around 5153\,\AA, blended with Cr\ii, Ti\ii, and Fe\i\ was used to derive the surface abundance of this
element, and we see it to vary between  $\log N_{\mathrm{Cu}}/N_{\mathrm{tot}}=-8.0$ and $-6.9$\,dex. At first glance,
it seems to be closely 
related to the distributions of Ni and Y, being enhanced at the magnetic
equator, but it exhibits in addition two less pronounced spots close to the magnetic poles.

\section{Effect of abundance inhomogeneities on spectral energy distribution} 
    
\label{models}
As already mentioned in Sect.\,\ref{intro} the light curve of \hd\ obtained by CoRoT 
has a periodic form with two clear maxima of slightly different atmplitudes at phases 
$\phi\,\approx\,0.05$ and $0.52$ (see Fig.~\ref{fig:phase_folded}).
This photometric variability is likely to be connected with inhomogeneous
surface element distributions, similar to those recently reported and 
successfully modeled by Krti\v{c}ka et al. (2007) for the hot CP2 star HD~37776.
The physical nature of this effect is directly connected with the
radiative flux redistribution due to enhanced or deficient opacity in the 
abundance spots relative to the rest of the stellar surface. Hence, as a star
rotates, the observer sees different stellar regions that are emitting a different amount
of radiative flux producing characteristic variability of indices in
phase-resolved photometry.

Of course, abundance spots are not the only effect which may produce visible rotational modulation in photometry. For instance, 
strong magnetic fields may also influence the total radiative flux gradients along the stellar surface since the magnetic
opacity and flux distribution strongly depend upon the modulus of the magnetic
field (Kochukhov~et~al.~2005, Khan \& Shulyak 2006). 
However, the mean surface magnetic field of \hd\ is too weak to noticeably
affect the integrated opacity coefficient, and thus the chemical spots are the most probable source of the observed variability. Details of our modelling of the
light variability taking into account surface abundance variations of \hd\ are
presented in this Sect.. 

Having produced abundance maps using the Doppler imaging technique it is, in
principle, possible to directly model the light curve
variability in the same manner as it was done in Krti\v{c}ka et al.
(2007). However, in case of 
HD~50773, there are several elements that play a noticeable role in the flux
redistribution,
which would require the calculation of hundreds of model atmospheres for an accurate surface 
integration.
These extensive computations are out of the scope of the present Paper.
Nevertheless, to explore the role of the mapped elements on the total energy balance in the
atmosphere of HD~50773 (and thus on the ability of these elements to affect the 
radiative balance) we computed a set of model atmospheres with different assumptions
about abundance patterns.
To carry out such calculations we employed the \llm\ stellar 
model atmosphere code (Shulyak et al. 2004), which incorporates treatment of individual
abundance patterns and computes the line opacity in a fine frequency grid for better integration of
the radiation field quantities.

First we computed the model with the surface averaged abundances of all nine elements
used in our DI analysis. Then another seven models were computed, individually increasing  
every element by $+2$~dex relative to its mean value. Once these computations were
done,
we integrated the output radiative fluxes from all the models and then compared
them to the flux produced by the mean abundance reference model. The flux integration was
performed in the wavelength range defined by the CoRoT photometric CCD, 
i.e. from $2500$\AA\ to $11000$\AA, with the resolution of $0.1$\AA, which is
the default flux resolution in \llm.  The effective temperature of all the
models was kept to be the same. 

We find that among the nine elements considered in this investigation 
only four have a clear influence on the model energy distribution by more than 1\%.
These elements are Cr (1.16\%), Fe (5.63\%),
Mg (3.02\%), and Si (2.21\%). The influence of the other five elements is much less pronounced
like, e.g. Ca (0.22\%).  
This agrees well 
with the recent study of Khan \& Shulyak (2007) who investigated the effect of individual chemistry on properties of CP stars
and showed the relative importance of Fe, Cr, and Si opacity for all the models considered.

\begin{figure*}
\includegraphics[width=0.5\textwidth,angle=-90]{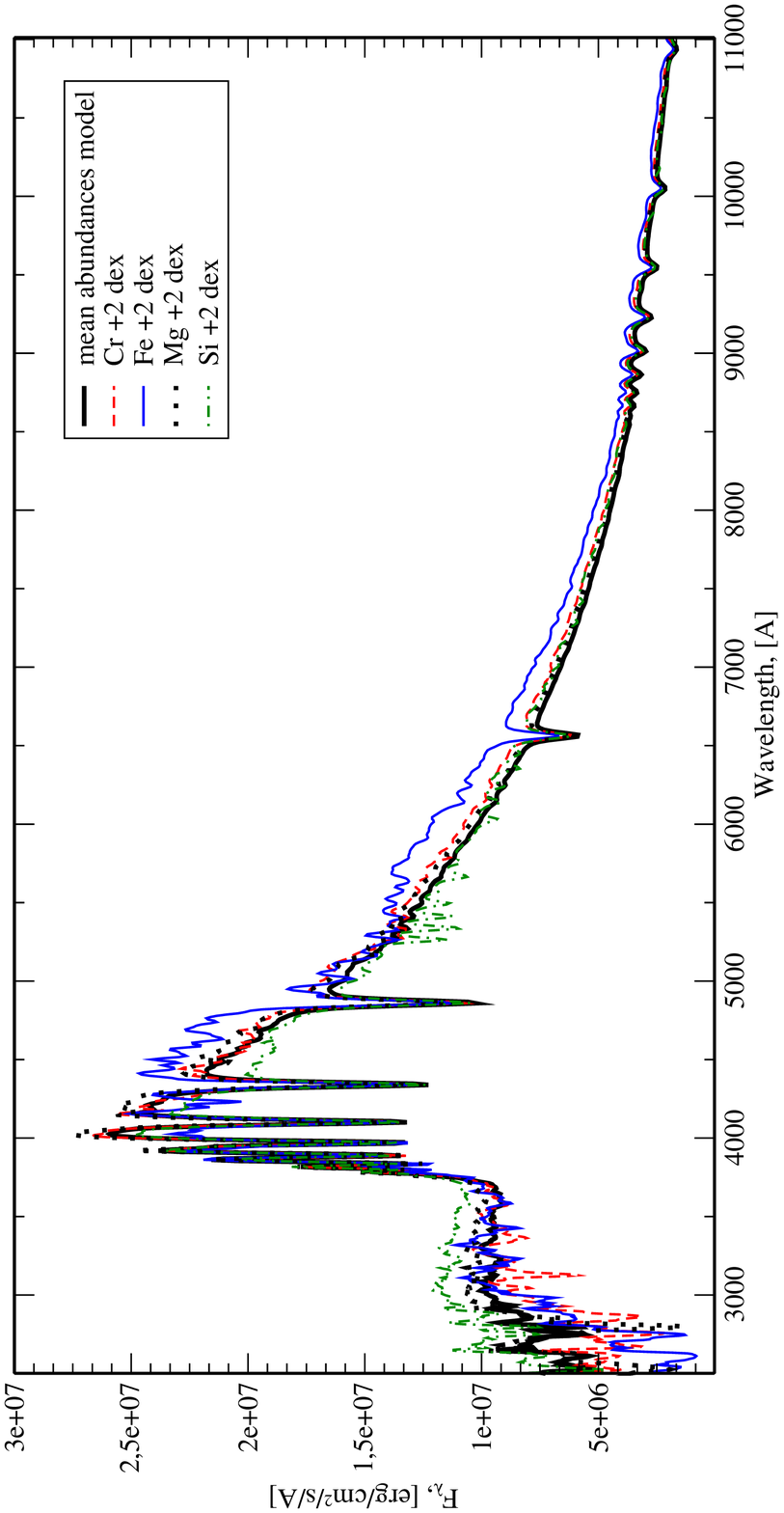}
\caption{The synthetic energy distributions of HD~50773 calculated with mean surface abundances and
abundances of Cr, Fe, Mg, and Si enhanced by $2$~dex. For all models $T_{\rm eff}=8300$\,K and 
$\log g=4.1$ were assumed. To provide a better view all fluxes were convolved with the resolution of $R=200$.}
\label{fig:sed}
\end{figure*}

As an example, Fig.\,\ref{fig:sed} demonstrates the synthetic energy distribution computed
with enhanced abundances of Cr, Fe, Mg, and Si. In particular, one can note the strong 
impact of Fe and Si on the shape of the energy distribution. However, the relatively small
changes in integrated flux values listed above are due to the fact that the CoRoT photometry
covers a wide wavelength region where the flux changes due to enhanced abundance
models which act in opposite directions in the UV and the visual regime. We do not show fluxes produced by
the other mapped elements since they lie extremely close to the flux of the mean abundance model (thick full line).

\section{Conclusions}             
\label{discuss}

The high-quality \co\ data enabled us by applying the Bayesian data analysis to derive stellar surface structures from
space quality photometry. Analysing an extensive set of ground based spectropolarimetric
data via Doppler imaging and Magnetic Doppler imaging made it possible to directly correlate 
the results from photometry to the chemical and magnetic stellar surface 
structure. 
The resulting abundances were used to model the light variability of \hd. 

The two areas covered by bright spots found with Bayesian data analysis of the
light curve coincide
very well with the magnetic polar regions on the surface of \hd.

The elemental abundance spots of the species contributing dominantly to the changes
in the integrated flux  
(Cr, Fe, and Si) are also clearly correlated with the magnetic field geometry
and hence with photometric spots.
Regions of overabundance are found close to
the magnetic poles and those of depletion linked to the magnetic equator. 
As described in Sect.\,\ref{models}, these elements produce bright spots in the photosphere.

Our results confirm the high potential of combining high quality photometric data
obtained in space with ground based spectroscopy. 
Further studies with a similar approach are currently ongoing and will
significantly  
increase the sample of stars analysed in this way, which is important for modelling elemental diffusion
in stars in the presence of a magnetic field.

\begin{acknowledgements}
We would like to thank the referee, Z. Mikul\'{a}\v{s}ek, for
very consctructive comments during the refereeing process.  
We also thank the \co, CFHT and TBL teams for their observing support and the
excellent data, and N.~Letourneur and J.--P.~Michel, who obtained the NARVAL spectra for us. 
This work was supported by the Austrian Science Fund FWF-P17580N2, the Lise
Meitner grant Nr. M998-N16 to DS and by the financial contributions 
of the Austrian Agency for International Cooperation in Education and Research (WTZ CZ-11/2008). 
OK is a Royal Swedish Academy of Sciences Research Fellow supported by a grant from the Knut and Alice Wallenberg
Foundation.
\end{acknowledgements}

\end{document}